# Low temperature ($T_s/T_m$ < 0.1) epitaxial growth of HfN/MgO(001) via reactive HiPIMS with metal-ion synchronized substrate bias

Michelle Marie S. Villamayor[a*], Julien Keraudy[a†], Tetsuhide Shimizu[b], Rommel Paulo B. Viloan[a], Robert Boyd[a], Daniel Lundin[c], J.E. Greene[a,d], Ivan Petrov[a,d], Ulf Helmersson[a]

[a] *Department of Physics, Linköping University, Linköping, SE-581 83, Sweden*

[b] *Division of Intelligent Mechanical Systems, Graduate School of System Design, Tokyo Metropolitan University, 6-6, Asahigaoka, Hino-shi, 191-0065 Tokyo, Japan*

[c] *Laboratoire de Physique des Gaz et Plasmas - LPGP, UMR 8578 CNRS, Université Paris-Sud, Université Paris-Saclay, F-91405 Orsay Cedex, France*

[d] *Frederick Seitz Materials Research Laboratory and Materials Science Department, University of Illinois, 104 South Goodwin, Urbana, Illinois 61801, USA*

## Abstract

Low-temperature epitaxial growth of refractory transition-metal nitride thin films by means of physical vapor deposition has been a recurring theme in advanced thin-film technology for several years. In the present study, 150-nm-thick epitaxial HfN layers are grown on MgO(001) by reactive high-impulse magnetron sputtering (HiPIMS) with no external substrate heating. Maximum film growth temperatures $T_s$ due to plasma heating range from 70-150 °C, corresponding to $T_s/T_m$ = 0.10-0.12 (in which $T_m$ is the HfN melting point in K). During HiPIMS, gas and sputtered-metal ion fluxes incident at the growing film surface are separated in time due to strong gas rarefaction and the transition to a metal-ion dominated plasma. In the present experiments, a negative bias of 100 V is applied to the substrate, either continuously during the entire deposition or synchronized with the metal-rich portion of the ion flux. Two different sputtering-gas mixtures, Ar/N$_2$ and Kr/N$_2$, are employed in order to probe effects associated with the noble-gas mass and ionization potential. The combination of x-ray diffraction, high-resolution reciprocal-lattice maps, and high-resolution

---

[*] Current affiliation: *Division of Solid-State Electronics, The Ångström Laboratory, Uppsala University, Box 534, SE-751 21 Uppsala, Sweden*

[†] Current affiliation: *Oerlikon Surface Solutions AG, Oerlikon Balzers, Iramali 18, LI-9496, Balzers, Liechtenstein*

cross-sectional transmission electron microscopy analyses establish that all HfN films have a cube-on-cube orientational relationship with the substrate, i.e., [001]$_{HfN}$||[001]$_{MgO}$ and (100)$_{HfN}$||(100)$_{MgO}$. Layers grown with continuous substrate bias, in either Ar/N$_2$ or Kr/N$_2$, exhibit a relatively high mosaicity and a high concentration of trapped inert gas. In distinct contrast, layers grown in Kr/N$_2$ with the substrate bias synchronized to the metal-ion-rich portion of HiPIMS pulses, have much lower mosaicity, no measurable inert-gas incorporation, and a hardness of 25.7 GPa, in good agreement with results for epitaxial HfN(001) layers grown at $T_s$ = 650 °C ($T_s/T_m$ = 0.26). The room-temperature film resistivity is 70 μΩ-cm, which is 3.2 to 10 times lower than reported values for polycrystalline HfN layers grown at $T_s$ = 400 °C.

## I. INTRODUCTION

Refractory transition-metal (TM) nitride thin films are employed in a wide variety of applications due to their unique combination of properties including high hardness,[1-5] scratch and abrasion resistance,[6] low coefficient of friction,[7] high-temperature oxidation resistance,[8-10] and tunable optical,[11,12] electrical[12-14] and thermal[15] properties. As a result, TM nitrides have gained considerable interest over the past decades and become technologically important for use as hard wear-resistant coatings,[16,17] decorative coatings,[18] and diffusion barriers;[15,19-24] the latter because of their high thermal stability[10,25] and low electrical resistivity.[13]

Epitaxial TM and rare-earth nitride (TiN,[26] CrN,[27] HfN,[28] CeN,[29] ZrN,[14] and VN[11]) layers have been grown on MgO(001) by reactive dc magnetically-unbalanced magnetron sputtering (DCMS) in either Ar/N$_2$ or pure N$_2$ environments at elevated temperatures $T_s$, typically between 600 and 850 °C, using low-energy, high-flux ion-irradiation of the film surface during growth. However, in many TM nitride applications, including diffusion barriers, there is a strong drive to grow dense high-crystalline-quality films at much lower temperatures in order to minimize deposition cycling times and allow the use of thermally-sensitive substrates. Thus, the quest for low-temperature epitaxy (LTE) of high-quality materials has been a recurring theme in materials science for several years with the expectation that the knowledge gained in such experiments will translate to the growth, at low temperatures, of dense, highly-oriented polycrystalline films.

Lee *et al*[30] demonstrated that single-crystal LTE TiN layers grown on MgO(001) substrates can be obtained in the absence of applied substrate heating using high-flux, low-energy ion irradiation during reactive DCMS in pure N$_2$. The maximum deposition temperature, due to plasma heating, was 420 °C ($T_s/T_m \simeq 0.22$, in which $T_m$ is the TiN melting point, 3220 K).[31] While $T_s$ was much less than for conventional epitaxial TiN(001), typically $\geq$ 700 °C ($T_s/T_m \simeq 0.30$),[26] it is still too high for many applications, including film growth on temperature-sensitive substrates such as polymers and light-weight metals (e.g., Li, Mg, and Al).

Here, we investigate the possibility of using reactive high-power impulse magnetron sputtering (HiPIMS) to grow epitaxial TM nitrides at much lower temperatures, $T_s/T_m \simeq 0.1$. HiPIMS allows ionization of up to 90% of sputtered metal atoms, depending upon pulsing conditions and the choice of metal target.[32] Thus, HiPIMS offers the opportunity to increase momentum transfer to the growing film, and enhance adatom mean-free paths due to irradiation by low-energy ionized sputtered *metal* atoms, rather than gas ions, during bias deposition.[33] Greczynski et al[34] probed the effect of metal- versus gas-ion irradiation during $Ti_{1-x}Al_xN$ film growth in mixed Ar/N$_2$ atmospheres using synchronized pulsed substrate bias in a hybrid HiPIMS/DCMS co-sputtering configuration. The results demonstrated that synchronizing the substrate bias with the metal-rich portion of HiPIMS pulses provides film densification, microstructure enhancement, surface smoothening, and decreased film stress with no measurable Ar incorporation.

We report the LTE growth of HfN on MgO(001), in the absence of applied substrate heating, using reactive HiPIMS in Ar/N$_2$ and Kr/N$_2$ discharges with low-energy ion irradiation of the growing film. The two gas mixtures are selected to provide significantly different momentum transfer from gas ions to the growing film ($m_{Ar}$ = 39.96 amu, $m_{Kr}$ = 83.79 amu, while $m_{Hf}$ = 178.49 amu), and different ionization potentials (the first ionization potential of Ar, $IP^1_{Ar}$, is 15.76 eV, while $IP^1_{Kr}$ = 14.00 eV).[31] HfN/MgO(001) is chosen as a model system for these experiments since: (1) HfN has the highest melting point among the TM nitrides ($T_m$ = 3310 °C [3583 K]),[31] (2) the HfN/MgO(001) system has a large lattice mismatch (7.46%),[28] and (3) the growth of high-quality HfN/MgO(001) single-crystals has been demonstrated at elevated temperatures.[28,35]

We isolate the effect of the nature of the incident ions (gas vs. metal) during film growth by comparing the results of two different modes of applied substrate bias: (i) continuously applied bias during the entire deposition and (ii) bias applied only in synchronous with the metal-rich portion of each HiPIMS pulse. The maximum substrate temperature, due to plasma heating during film growth, is 150 °C ($T_s/T_m$ < 0.12) with continuously-applied bias and 70 °C ($T_s/T_m$ < 0.10) using synchronized bias.

X-ray diffraction (XRD) $\omega$-$2\theta$ and azimuthal $\varphi$ scans, high-resolution reciprocal-lattice maps (HR-RLMs), and high-resolution cross-sectional transmission electron microscopy (XTEM) analyses establish that all HfN films have a cube-on-cube relationship with their MgO(001) substrates. The epitaxial layers with the highest quality, grown with synchronized bias in Kr/N$_2$ discharges, are essentially fully-relaxed with no measurable concentration of trapped gas atoms and, hence, low film stress. They also have much lower room-temperature resistivity, 70 vs. 100-130 $\mu\Omega$-cm, and a slightly lower hardness, 25.7 vs. 26.8-27.6 GPa, than compressively-stressed layers grown with continuous bias. A comparison of LTE HfN(001) films grown with a continuous substrate bias in Ar/N$_2$ vs. Kr/N$_2$ discharges reveals that the layers deposited in Kr/N$_2$ exhibit lower mosaicity.

## II. EXPERIMENTAL DETAILS

All HfN layers, 150-nm thick, are grown by magnetically-unbalanced reactive HiPIMS in mixed noble-gas/N$_2$ discharges in a load-locked stainless-steel UHV system with a base pressure of $1\times10^{-9}$ Torr ($1.3\times10^{-7}$ Pa). The target is a Hf disk (99.9% purity) with a diameter of 75 mm and a thickness of 6 mm. The N$_2$ (99.9995% pure) flow rate is maintained constant at 2.0 sccm, while the noble gas, Ar or Kr (both 99.9997% pure), is introduced through a leak valve to maintain a constant pressure of 3 mTorr (0.4 Pa) with a N$_2$ fraction of 12 mol%.

Polished MgO(001) wafers, $10\times10\times0.5$ mm$^3$, whose quality is confirmed following the XRD procedure of Schroeder *et al*,[36] are used as substrates. The wafers are cleaned using successive rinses in ultrasonic baths of acetone, isopropanol, and de-ionized water. They are then blown dry with N$_2$, mounted on a Mo substrate holder, and inserted into a differentially-pumped load-lock chamber for transport to the deposition chamber where the wafers are thermally degassed at 800 °C for 1h, a procedure shown to result in sharp MgO(001)1×1 reflection high-energy electron diffraction patterns.[37] Following degassing, the substrates are allowed to cool to room temperature. The target-to-substrate separation is 11 cm.

Prior to initiating deposition, the Hf target is sputter etched at 50 W for five minutes in either pure Ar, or pure Kr, discharges at 3 mTorr with a shutter shielding the substrate. During film growth, no external substrate heating is applied. The film-growth temperature $T_s(t)$, due to plasma heating, is measured as a function of deposition time $t$ using a thermocouple attached to the top of a dummy HfN-coated MgO substrate.

HiPIMS pulses, 100 μs in length at a frequency of 100 Hz (duty cycle = 1%), are supplied by a HiPSTER 1 pulsing unit (Ionautics AB) fed by a dc power supply. The target current and voltage peak shapes are recorded and monitored with a Tektronix TDS 520C digital oscilloscope connected directly to the HiPIMS pulsing unit. Sputter deposition is carried out at a constant average target power of ~150 W. A negative substrate bias $V_s$ = 100 V is applied using one of two biasing schemes: (i) continuous bias during the entire HiPIMS pulse, or (ii) bias synchronized, using a second HiPSTER 1 unit, to the metal-rich portion of the discharge, as determined by mass spectrometry, during each HiPIMS pulse. For films grown with synchronized bias, ions arrive at the substrate during the remainder of the pulse with an energy $E_s = -e(V_p - V_f) \approx 10$ eV ($V_p$ is the plasma potential and $V_f$ is the substrate floating potential), which is below the bulk lattice-atom displacement threshold for TM nitrides.[38]

*In situ* time-dependent mass and energy spectroscopy analyses of ion fluxes incident at the substrate plane are carried out using a Hiden Analytical PSM003 instrument to determine relative compositions, charge states, and energies as a function of the choice of inert gas and bias scheme. The mass-spectrometer orifice is electrically grounded during these experiments and the ion energy is scanned from 1 to 80 eV in 0.5 eV steps. Further details regarding the mass spectrometer can be found elsewhere.[39] The separation between the mass spectrometer orifice, located below the substrate position, and the center of the target is 21 cm. Thus, measured energy distributions underestimate ion-impact energies at the substrate due to additional gas-phase collisions occurring between the substrate plane position and the entrance to the mass spectrometer. Ion times-of-flight (TOF) are corrected, following the procedure of Bohlmark *et al*,[39] for the difference in position and listed in Table 1 for the lowest and highest incident kinetic energies.

Ion-energy-distribution functions (IEDFs) are recorded for $Hf^+$, $Hf^{2+}$, $Ar^+$, $Ar^{2+}$, $Kr^+$, $Kr^{2+}$, $N^+$, and $N^{2+}$ ions, while sputtering under the same conditions in either $Ar/N_2$ or $Kr/N_2$ gas mixtures. Since the quadrupole mass analyzer has a higher transmission at lower mass,[40] and because of the large mass differences between metallic and gaseous ions in these investigations (ranging from 14 to 180 amu, see Table 1), the spectrometer settings are separately tuned for each species.

**Table 1.** Ion species, mass/charge (m/z) ratios, incident ion-kinetic-energy ranges, corrected TOF values, and energy-averaged TOF values.

| Ion | m/z | Kinetic-energy range [eV] | TOF [μs] | Average TOF [μs] |
|---|---|---|---|---|
| Hf$^+$ | 180 | 1-80 | 160-153 | 156.5 |
| Hf$^{2+}$ | 90 | 1-80 | 113-108 | 110.5 |
| Ar$^+$ | 40 | 1-80 | 76-73 | 74.5 |
| Ar$^{2+}$ | 20 | 1-80 | 54-52 | 53.0 |
| Kr$^+$ | 84 | 1-80 | 81-77 | 106.5 |
| Kr$^{2+}$ | 42 | 1-80 | 48-46 | 79.0 |
| N$_2^+$ | 28 | 1-80 | 66-63 | 64.5 |
| N$^+$ | 14 | 1-80 | 48-46 | 47.0 |

Time-resolved measurements of ion-flux energy distributions are collected during 50 consecutive pulses such that the total acquisition time per data point is 2 ms. To synchronize the measurements with the target-pulse onset, the mass-spectrometer circuit is triggered by a transistor-transistor logic (TTL) pulse sequence generated at the output of the HiPSTER synchronization unit. The gate width is 5 μs, with a 1 μs delay following the pulse onset.

Compositions of as-deposited HfN$_x$ layers are determined by time-of-flight elastic recoil detection analyses (TOF-ERDA) employing a 36 MeV $^{127}$I$^{8+}$ probe beam incident at 67.5° with respect to the sample normal; recoils are detected at an angle of 45°. The results are analyzed using CONTES software.[41] Uncertainties in reported $x$ values for HfN$_x$ films are less than ±0.025.

A high-resolution PANalytical X'pert XRD diffractometer with a CuKα$_1$ source ($\lambda$ = 1.540597 Å) and a four-crystal Ge(220) monochromator is used for determining film orientation. Lattice parameters perpendicular $a_\perp$ and parallel $a_\parallel$ to the film surface, and residual strains $\varepsilon$, are obtained from high-resolution reciprocal lattice maps (HR-RLMs) around asymmetric 113 reflections. The RLMs consist of a series of $\omega$-$2\theta$ scans acquired over a range of ω offsets. High-resolution transmission electron microscopy (HR-TEM) analyses are carried out in a FEI Tecnai G$^2$ TF 20 UT instrument operated at 200 kV. Cross-sectional specimens are prepared using a two-step procedure consisting of mechanical polishing followed by Ar$^+$-ion milling at a shallow incidence angle of 6° from the sample surface, with gradually decreasing ion energies: 5 keV initially, followed by 2.5, and 1 keV.

Room-temperature electrical resistivities $\rho$, as a function of film-growth conditions, are obtained from four-point probe results, with each reported $\rho$ value averaged over ten separate measurements. Nanoindentation analyses of as-deposited HfN films are performed using a Hysitron TI-950 Triboindenter equipped with a sharp Berkovich diamond probe calibrated using a fused-silica standard and a single-crystal TiN(001) reference sample.[28] A minimum of 30 indents is made in each specimen with maximum loads of 1.5 mN. Indentation depths are never allowed to exceed 10% of the film thickness in order to minimize substrate effects. Film hardnesses $H$ and indentation moduli $E$ are determined using software based on the Oliver and Pharr method.[42]

Transport of ions in matter (TRIM),[43] a Monte Carlo code included in the stopping power and range of ions in matter (SRIM) software package, is used to estimate the reflection probability and average energy of noble-gas (Ar or Kr) atoms backscattered from a HfN target, as well as primary-ion and recoil projected ranges of metal- and gas-ions incident at the growing film during deposition.

## III. RESULTS

### A. Plasma characterization

Prior to initiating film-growth experiments, reactive-HiPIMS discharges in mixed Ar/N$_2$ and Kr/N$_2$ environments are characterized in order to design substrate bias strategies as described in subsections III.B and III.C. Figure 1 presents typical target current $I_T(t)$ and voltage $V_T(t)$ waveforms acquired during HIPIMS sputtering of Hf in Ar/N$_2$ and Kr/N$_2$ atmospheres. Both voltage waveforms are rectangular in shape with the maximum Kr/N$_2$-discharge voltage 100 V higher than that of Ar/N$_2$.

$I_T(t)$ waveforms exhibit delays in the target current rise onset of ~10 μs for Kr/N$_2$ and ~25 μs for Ar/N$_2$, after which the current increases rapidly to reach a maximum value at 50-60 μs, and then decreases to a saturation value at ~90 μs, before falling to zero at the end of the HiPIMS pulse. The reason for the earlier current increase in the Kr/N$_2$ discharge during pulse onset is due to a significantly higher Kr vs. Ar electron-impact-ionization cross section,[44] which decreases the time required to accumulate a sufficiently high electron density for gas breakdown (i.e., rapid discharge current increase). However, the differences in $I_T(t)$ during the remainder of the discharge pulse arise due to significant recycling of the ionized process gas (primarily Ar or Kr),[45,46] since the sputter yield of Hf by $Ar^+$ ions ($S_{Hf}^{Ar^+} \simeq 0.4$) and by $Kr^+$ ions ($S_{Hf}^{Kr^+} \simeq 0.8$) are well below unity, thus limiting self-sputter recycling.[46] A slightly lower maximum discharge current is expected for Kr/$N_2$ discharges, since the ionization of Kr required for gas recycling is reduced compared to Ar by the lower electron temperature arising due to: (1) a higher concentration of Hf (for which $S_{Hf}^{Ar^+} > S_{Hf}^{Kr^+}$) available in the ionization region with a much lower first-ionization potential, $IP_{Hf^+}^1 =$ 6.83 eV compared to the process gas,[46] and (2) $Kr^+$ having a lower secondary-electron-emission yield than $Ar^+$.[47]

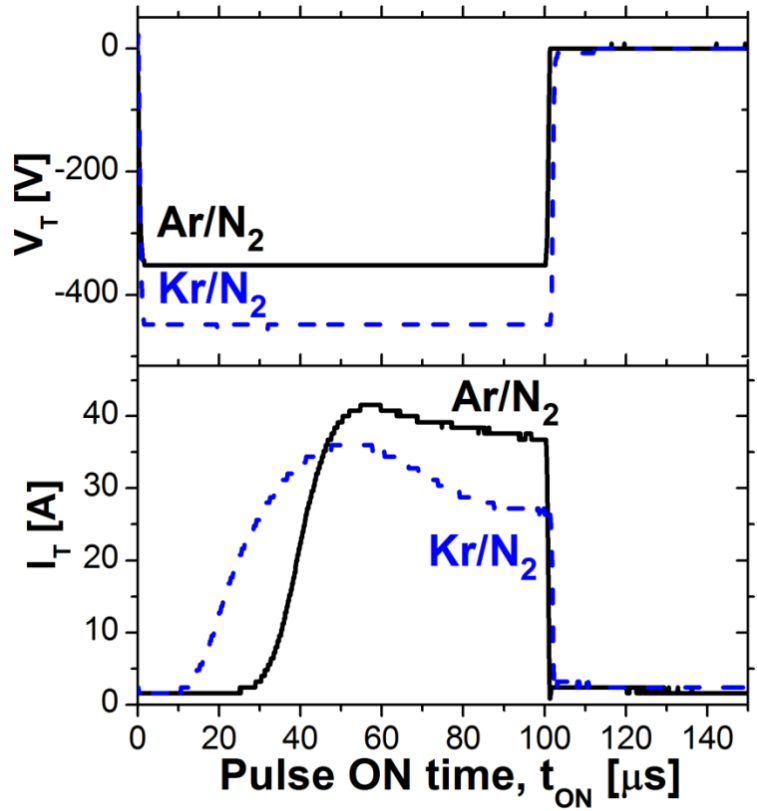

**Figure 1.** Target voltage $V_T$ (upper panel) and current $I_T$ (lower panel) waveforms during reactive HiPIMS sputtering of Hf in Ar/$N_2$ (solid black line) and Kr/$N_2$ (dashed blue line) discharges at 3 mTorr.

*In situ* time-averaged IEDFs for $Ar^+$, $N^+$, $N_2^+$, $Hf^+$, and $Hf^{2+}$ gas and metal ions incident at the substrate plane during reactive HiPIMS sputtering of Hf in Ar/$N_2$ and Kr/$N_2$ atmospheres are shown in Figures 2(a) and 2(b), respectively. $Ar^{2+}$ and $Kr^{2+}$ ions are also detected, but not shown here since the sum of their contribution to the total measured ion intensity is < 1%. All gas- and metal-ion IEDFs exhibit a narrow low-energy peak located at 1.4 eV in Ar/$N_2$ and 2.0 eV in Kr/$N_2$ discharges. These values are consistent with previous mass spectrometry investigations of HiPIMS discharges,[48-50] and represent the potential drop between the bulk plasma potential and the grounded orifice of the mass spectrometer, across which thermalized ions are accelerated.[48] $Hf^+$ and $Hf^{2+}$ metal-ion IEDFs, for both $Ar/N_2$ and $Kr/N_2$ discharges, exhibit an intense broad peak with a maximum near 10 eV, followed by a high-energy tail extending to 80 eV, which we attribute to a superposition of a Sigmund-Thompson sputtered-species energy distribution,[51,52] ion acceleration by the combination of plasma and floating potentials, and HiPIMS plasma instabilities due to collective plasma effects.[53-55]

$Ar/N_2$ and $Kr/N_2$ gas-ion IEDFs at the substrate plane display significant differences. As depicted in Figure 2(a), $Ar^+$ IEDFs possess a broad high-energy peak centered at ~65 eV. This differs dramatically from $Kr^+$ IEDFs (a typical result is shown in Figure 2(b)), for which only the low-energy peak is observed. The high-energy $Ar^+$ ions observed in $Ar/N_2$ HiPIMS discharges arise primarily from the high probability of $Ar^+$ ions incident at the target being neutralized and reflected from heavy Hf target atoms toward the substrate plane with a significant fraction of the incident energy retained.[56,57]

An estimate of the *maximum* backscattering energy $E_{b,Ar}$ of $Ar^+$ ($m_{Ar}$ = 39.95 amu) ions in 180° reflections from Hf ($m_{Hf}$ = 178.49 amu) target atoms is given by the expression[57,58]

$$E_{b,Ar} = \left(\frac{m_{Hf} - m_{Ar}}{m_{Hf} + m_{Ar}}\right)^2 \cdot E_i, \qquad (1)$$

for which $E_i$ is the kinetic energy of the incident $Ar^+$ ion, 350 eV, corresponding to acceleration across the cathode sheath. Equation (1) yields a maximum Ar backscattering energy $E_{b,Ar}$ = 140 eV. The corresponding value for $Kr^+$ is $E_{b,Kr}$ = 59 eV, obtained using the higher incident ion energy of 450 eV shown in Fig. 1. TRIM[43] simulations yield reflection probabilities of 0.27 and 0.10 and reflected energies, averaged over all backscattering angles, of 85 and 39 eV for Ar and Kr, respectively. The angular scattering distribution from the target, combined with the distribution of ionization mean free paths, accounts for the large energy spread in recorded $Ar^+$ IEDFs.

$N_2^+$ IEDFs exhibit only intense low-energy peaks, with no high-energy peaks or tails. A similar behavior was recently reported by Greczynski *et al* during reactive HiPIMS sputtering of Hf in $Ar/N_2$.[58] $N_2^+$ ions, upon impingement at the Hf target, are dissociatively neutralized to two N atoms, which can be backscattered as fast N atoms since $m_N \ll m_{Hf}$. These fast backscattered N atoms are partially responsible for the pronounced high-energy tails in $N^+$ IEDFs from both Ar/N$_2$ and Kr/N$_2$ discharges. The fact that the $N^+$ IEDFs closely resemble the corresponding $Hf^+$ and $Hf^{2+}$ IEDFs suggests that a significant fraction of the energetic $N^+$ ions also originate from sputter-ejected N atoms. Modeling studies of reactive HiPIMS discharges have confirmed the importance of the target as a source of reactive gas atoms in Ar/O$_2$ discharges,[59] and we expect a similar effect for Ar/N$_2$.

The gas-ion IEDF intensities in Figure 2 are significantly higher in Ar/N$_2$ than in Kr/N$_2$ discharges. In the former case, the total energy-integrated ion intensity $J$, with energies above 5 eV, is dominated by gas ions ($J_{gas}/J_{metal} = 2.78$); while for Kr/N$_2$ discharges, $J$ is controlled by $Hf^+$ and $Hf^{2+}$ metal ions ($J_{gas}/J_{metal} = 0.54$). Furthermore, the doubly-to-singly charged Hf ion intensity ratio is significantly higher in Ar/N$_2$, $J_{Hf^{2+}}/J_{Hf^+} = 1.7$, than in Kr/N$_2$ HiPIMS discharges, $J_{Hf^{2+}}/J_{Hf^+} = 0.3$. This is consistent with previous results for non-reactive HiPIMS sputtering of Ti in Ar/N$_2$ versus Kr/N$_2$ discharges.[60]

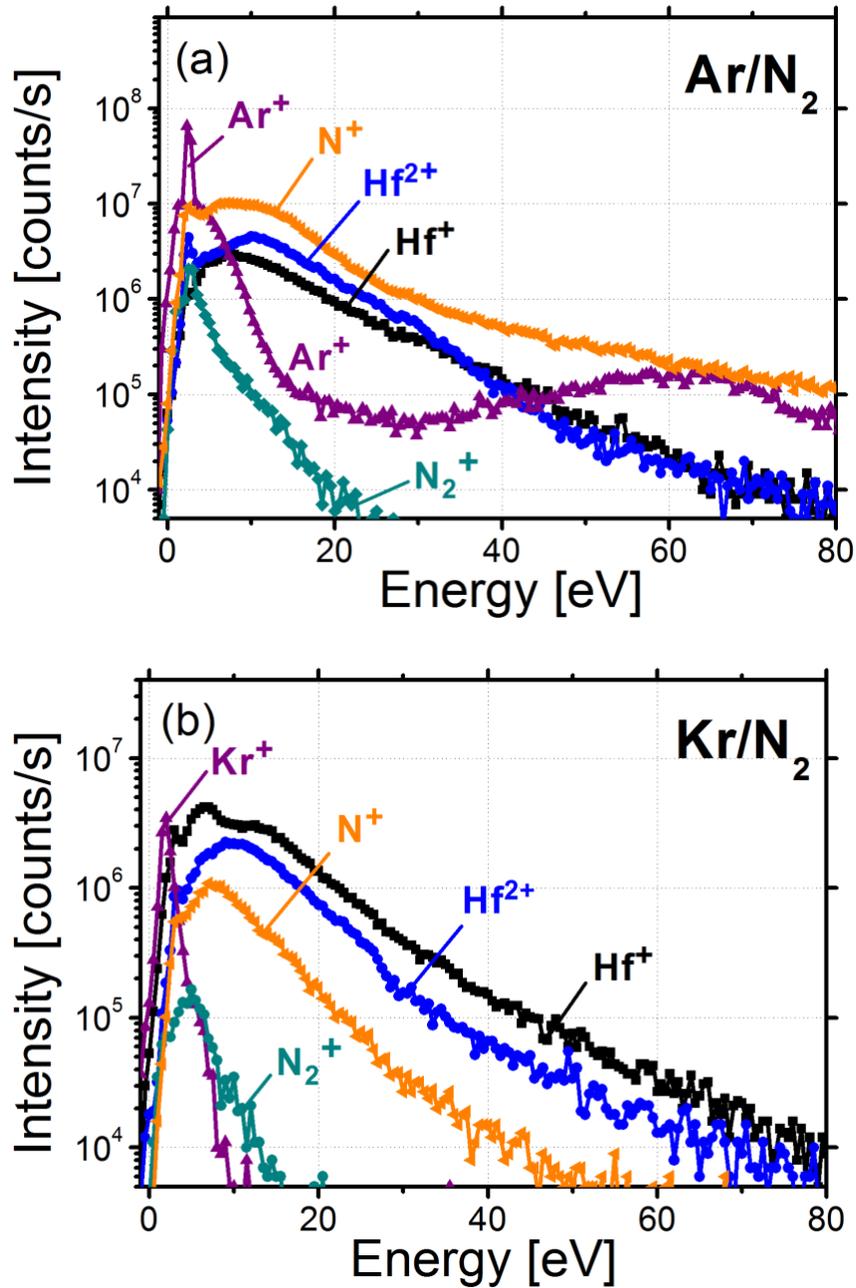

**Figure 2.** Time-averaged IEDFs for ions during reactive HiPIMS sputtering of Hf in (a) Ar/N$_2$ and (b) Kr/N$_2$ discharges at 3 mTorr.

Time-resolved ion intensities incident at the substrate plane during HiPIMS pulses in Ar/N$_2$ and Kr/N$_2$ discharges are displayed in Figures 3(a) and 3(b). Data points are obtained every 10 µs by integrating over the energy range from 0 to 80 eV. The non-zero intensities detected at $t = 0$ (start of the voltage pulse) in both Figures are due to residual ionized species from the previous pulse, as reported for reactive HiPIMS sputtering of Cr in Ar/N$_2$ mixtures.[61] With both Ar/N$_2$ and Kr/N$_2$ discharges, the total ion intensity during the early part of the HiPIMS pulses (Figure 3) is dominated by noble-gas ions. At 40 μs into the HiPIMS pulse, N$^+$ ions represent ~50% of the total measured ion intensity incident at the substrate in Ar/N$_2$ and ~30% in Kr/N$_2$.

Hf$^+$ and Hf$^{2+}$ ion intensities begin to increase rapidly as gas-ion intensities decrease at $t \sim 50$ μs, in both Ar/N$_2$ and Kr/N$_2$ discharges, due to a strong gas rarefaction and the increased availability of Hf neutrals that leads to a decrease in the average plasma electron temperature. The latter effect is the result of Hf having a lower first ionization potential than those of the two noble gases and nitrogen (N$_2$ and N) as shown in Table 2. Figure 3(a) reveals that measured gas- and metal-ion intensities are of comparable values at $t$ between ~100 and 170 μs in Ar/N$_2$ HiPIMS discharges; no clear metal-ion dominated mode is established. At $t = 120$ μs, $J_{\text{gas}}/J_{\text{metal}}$ reaches a minimum value of ~1.8, with Hf$^{2+}$ ions contributing the major fraction, accounting for ~65% of the total metal-ion intensity. In distinct contrast, for Kr/N$_2$ discharges, the ion flux shifts from being gas-dominated to metal-dominated at $t \sim 55$ μs (Figure 3(b)). At 100 μs, metal ions (Hf$^+$ + Hf$^{2+}$) represent ~90% of the total ion intensity with the major contribution, ~70%, being singly-charged Hf$^+$ ions. At longer times, $t > 200$ μs (100 μs after pulse termination), gas rarefaction has decreased due to gas refill and noble-gas ions again dominate the total ion flux in both Ar/N$_2$ and Kr/N$_2$ discharges.

**Table 2.** The first and second ionization potentials, $IP^1$ and $IP^2$, for metal and gas species used in these experiments.[31]

| Atom or molecule | $IP^1$ [eV] | $IP^2$ [eV] |
|---|---|---|
| **Hf** | 6.83 | 15 |
| **Ar** | 15.76 | 27.63 |
| **Kr** | 14.00 | 24.36 |
| **N** | 14.53 | 29.60 |
| **N$_2$** | 15.58 | |

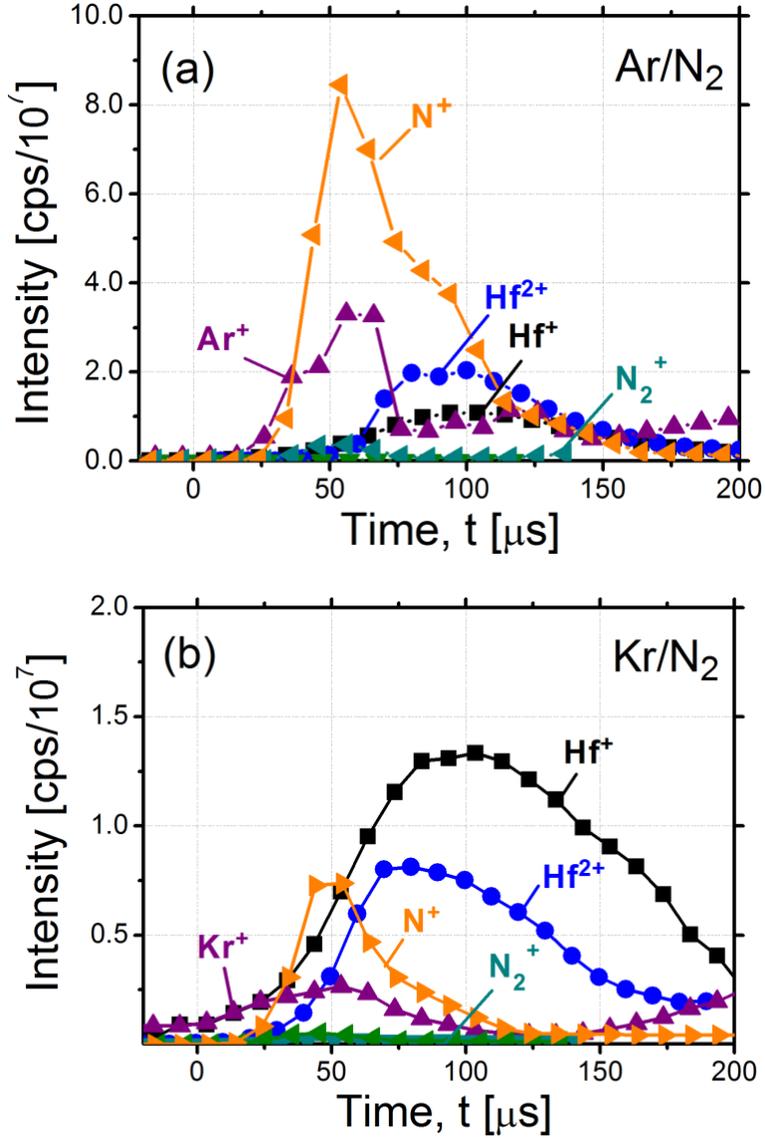

**Figure 3**. Time evolution of energy-integrated ion intensities incident at the substrate plane during reactive-HiPIMS sputtering of Hf in: (a) Ar/N$_2$ and (b) Kr/N$_2$ discharges at 3 mTorr.

## B. HfN film growth

Based upon the plasma characterization results in Section III.A, three sets of HfN films are deposited using two different substrate-bias modes with $V_s$ = 100 V. In mode (i), $V_s$ is applied continuously throughout film deposition in order to investigate the effects of gas-ion irradiation during reactive HfN film growth in Ar/N$_2$ vs. Kr/N$_2$ discharges; while in mode (ii), $V_s$ is applied only in synchronous with the metal-rich portion of each Kr/N$_2$ HiPIMS pulse, from $t$ = 60 to 160 μs (i.e., 60 μs after pulse termination). We did not carry out synchronized-bias experiments in Ar/N$_2$ discharges since time-resolved mass-spectrometry measurements provided no evidence for the presence of a clear metal-ion-dominated regime.

In bias mode (i) with $V_s$ applied continuously, the film-growth temperature $T_s(t)$ increases approximately linearly as a function of deposition time from room temperature to a saturation value of 150 °C ($T_s/T_m < 0.12$) after ~30 min during 60-minute deposition runs in both Ar/N$_2$ and Kr/N$_2$ discharges. In Kr/N$_2$ discharges with a synchronous bias, mode (ii), $T_s$ increases to a saturation value of 70 °C ($T_s/T_m < 0.10$) 40 mins into 60-minute deposition runs.

ERDA results show that HfN films grown under continuous substrate bias in Ar/N$_2$ are overstoichiometric, with a N/Hf ratio of 1.15, and O and Ar concentrations of 3.6 and 4.0 at%, respectively. Switching the inert gas from Ar to Kr, while maintaining the substrate bias in a continuous mode, decreases the N/Hf ratio to 1.10, with O and Kr concentrations of 1.8 and 3.0 at%. For HfN films grown with synchronized bias, the layers are near-stoichiometric with N/Hf = 1.05 and an O concentration of 1.8 at%. The trapped Kr concentration is ≲ 0.6 at.%.

### C. HfN film nanostructures

Figure 4 consists of typical $\omega$-$2\theta$ XRD scans from HfN films grown with both continuous and synchronous substrate bias. Each scan contains two distinct peaks centered at 39.75±0.03° and 43.0°±0.02°, which are attributed to 002 reflections from HfN and MgO, respectively.[28] Both the K$\alpha_1$ and K$\alpha_2$ peaks of the MgO 002 reflection are resolved. The XRD scans reveal changes in the 002 HfN peak intensity $I_{002}$, the full width at half maximum (FWHM) intensity $\Gamma_{2\theta}$, and peak positions depending on choice of substrate bias mode and noble gas. For samples deposited with continuous bias, $I_{002}$ increases from 7.6×10$^3$ cps with Ar/N$_2$ to 9.6×10$^3$ cps with Kr/N$_2$, while FWHM values $\Gamma_{2\theta}$ decrease from 0.85° to 0.70°. The results indicate enhanced crystalline quality for HfN films grown in Kr/N$_2$ discharges. In addition, the peak position shifts to slightly lower $2\theta$ values, from 39.78° to 39.75°, as the noble-gas constituent is switched from Ar to Kr. For comparison, a $2\theta$ value of 39.721° was reported for epitaxial stoichiometric HfN/MgO(001) grown at $T_s$ = 650 °C, and $2\theta$ was found to range from 39.6° to 39.8° for HfN$_x$ as $x$ was increased from 0.85 to 1.2.[28,35]

Synchronizing the substrate bias to the metal-rich portion of the HiPIMS pulses during HfN film growth in Kr/N$_2$ yields 002 peak intensities, $I_{002}$ = 8.7×10$^4$ cps, which are approximately an order of magnitude higher than those obtained with a continuous substrate bias in Kr/N$_2$. In addition, the 002 peak FWHM intensity $\Gamma_{2\theta}$ decreases by more than a factor of 2.3× to 0.30°. The synchronized-bias 002 peak position is 39.73°.

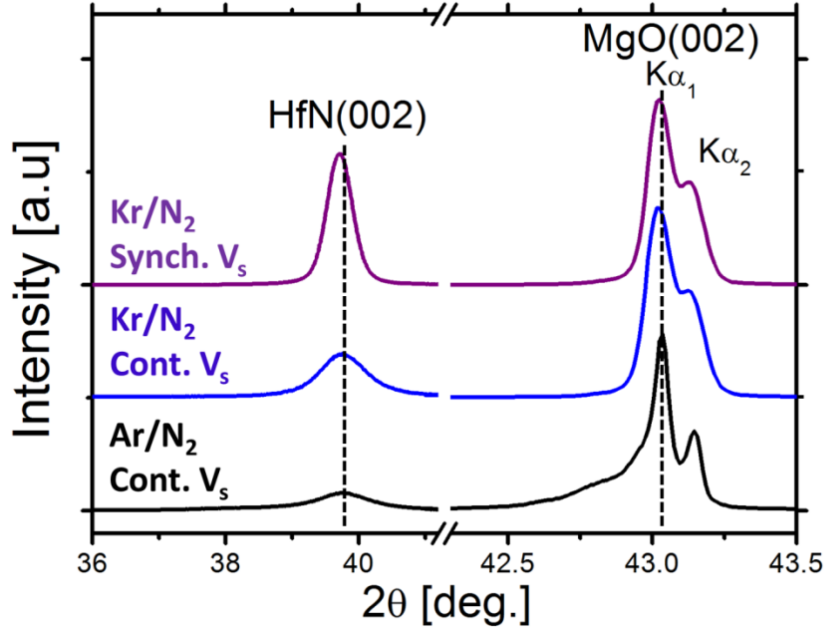

**Figure 4**. Typical $\omega$-$2\theta$ XRD scans (displaced vertically for convenience in viewing) from 150-nm-thick HfN layers grown on MgO(001) by reactive HiPIMS. The lower two scans are from films deposited in Ar/N$_2$ and Kr/N$_2$ discharges at 3 mTorr with a negative substrate bias, $V_s$ = 100 V, applied continuously during HiPIMS pulses. The upper scan is from a film grown in Kr/N$_2$ with the substrate bias synchronized to the metal-rich portion of each pulse. The average target power in all cases is 150 W.

$\omega$-rocking curves, for the same three samples corresponding to the $\omega$-$2\theta$ scans in Figure 4, are displayed in Figure 5. The HfN$_x$ layer grown with a synchronized bias in Kr/N$_2$ exhibits the highest 002 HfN peak intensity and a much lower FWHM $\Gamma_\omega$ value, 1.04°. FWHM $\Gamma_\omega$ values for films grown with a continuous substrate bias are 4.70° (Kr/N$_2$) and 5.00° (Ar/N$_2$).

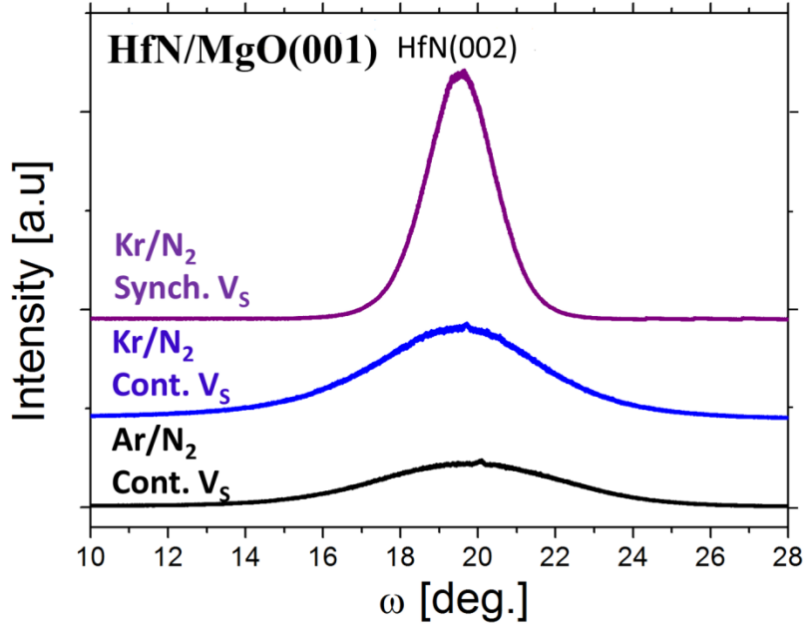

**Figure 5.** Typical 002 ω-rocking curves from 150-nm-thick HfN layers grown on MgO(001) by reactive HiPIMS deposition. The lower two scans are from films deposited in mixed Ar/N₂ and Kr/N₂ discharges at 3 mTorr with a negative substrate bias $V_s$ = 100 V applied continuously during HiPIMS pulses. The upper scan is from a film grown in Kr/N₂ with the substrate bias synchronized to the metal-rich portion of the pulses. The average target power in all cases is 150 W.

HfN(001) in-plane $\xi_{\parallel}$ and out-of-plane $\xi_{\perp}$ x-ray coherence lengths, which are directly related to the sample mosaicity, a measure of crystalline quality, are obtained from the widths of the 002 diffracted intensity distributions perpendicular $\Delta k_{\perp}$ and parallel $\Delta k_{\parallel}$ to the diffraction vector using the relationships[37]

$$\xi_{\parallel} = \frac{2\pi}{|\Delta k_{\perp}|} = \frac{\lambda}{2\Gamma_{\omega}\sin\theta} \qquad (2)$$

and

$$\xi_{\perp} = \frac{2\pi}{|\Delta k|} = \frac{\lambda}{\Gamma_{2\theta}\cos\theta}, \qquad (3)$$

for which $\Gamma_{2\theta}$ and $\Gamma_{\omega}$ are the FWHM intensities, after correction for instrumental broadening, of the 002 Bragg peak in the ω and 2θ directions, respectively, and θ is the Bragg angle. Using the data presented in Figures 4 and 5, the coherence lengths for HfN layers deposited in the two noble gas mixtures with continuous substrate bias are similar: $\xi_{\parallel} = 26$ Å and $\xi_{\perp} = 114$ Å for films grown in Ar/N$_2$, and $\xi_{\parallel} = 28$ Å with $\xi_{\perp} = 134$ Å in Kr/N$_2$. For HfN layers grown in Kr/N$_2$ with a synchronized bias, the coherence lengths are significantly larger: $\xi_{\parallel} = 93$ Å and $\xi_{\perp} = 313$ Å. Thus, the use of primarily metal-ion, rather than gas-ion, irradiation during the growth of HfN(001) results in films with much lower mosaicity and higher crystalline quality.

Figure 6 shows typical HfN/MgO(001) HR-XRD $\phi$ scans, obtained in parallel-beam mode, with $\omega$ and $2\theta$ angles set to detect 220 peaks at a tilt angle of 45° with respect to the surface normal, from layers grown in Kr/N$_2$ with synchronized bias. The $\phi$ scans exhibit four 90°-rotated 220 peaks at angular positions which are identical for HfN and the MgO substrate. Similar results are observed for samples deposited with continuous bias. The combination of the XRD $\omega$-$2\theta$ and $\phi$ scans demonstrate that all HfN(001) layers grow epitaxially with a cube-on-cube relationship to the substrate: (001)$_{HfN}$∥(001)$_{MgO}$ and [100]$_{HfN}$∥[100]$_{MgO}$.

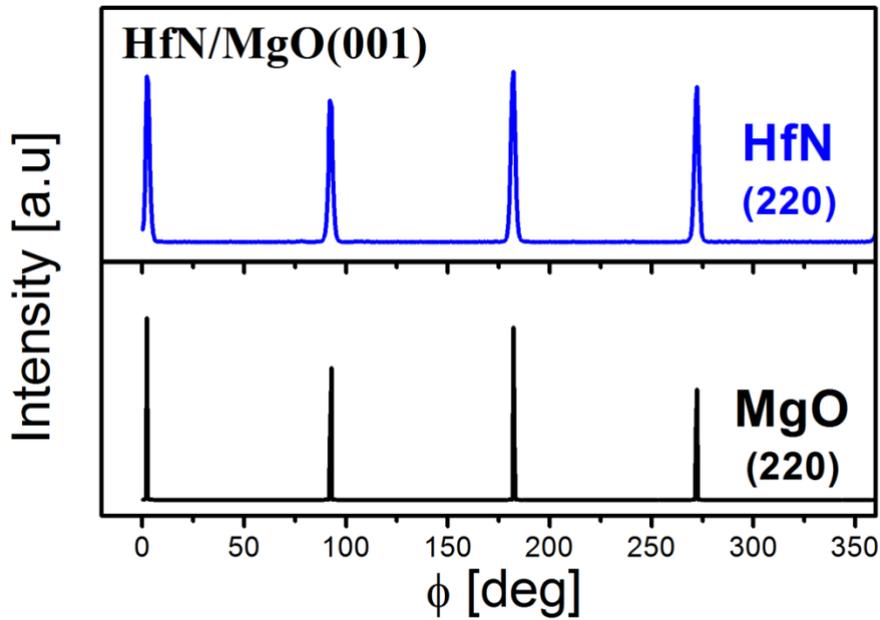

**Figure 6**. (Upper panel) Typical XRD 220 $\phi$-scan from a 150-nm-thick HfN/MgO(001) layers grown by reactive HiPIMS in Kr/N$_2$ discharges with $V_s = 100$ V synchronized to the metal-rich portion of each pulse. The average target power is 150 W at a pressure of 3 mTorr. (Lower panel) XRD 220 $\phi$-scan from a bare MgO substrate.

Relaxed lattice constants and residual strains for HfN layers grown in Kr/N$_2$ discharges under continuous and synchronized bias are determined from analyses of HR-RLM results. Figure 7 consists of typical HR-RLMs acquired about the asymmetric (113) reflection from HfN/MgO(001) layers deposited in Kr/N$_2$ with continuous bias (Figure 7(a)) and synchronized bias (Figure 7(b)). The dashed lines in the two panels extend from the origin, located outside the panels, along the 113 direction. Diffracted intensity distributions are plotted as isointensity contours as a function of the reciprocal lattice vectors $k_\parallel$ parallel and $k_\perp$ perpendicular to the surface. $k_\parallel$ and $k_\perp$ are related to $\theta$ and $\omega$ peak positions via the equations[62]

$$k_\parallel = 2r_E \sin(\theta) \cos(\omega - \theta) \tag{4}$$

and

$$k_\perp = 2r_E \sin(\theta) \sin(\omega - \theta), \tag{5}$$

in which $r_E$ is the Ewald sphere radius, given by $r_E = 1/\lambda$. For a 113 reflection from a 001-oriented NaCl-structure sample, the in-plane $a_\parallel$ and out-of-plane $a_\perp$ lattice parameters are obtained from the relationships $a_\parallel = \sqrt{2}/k_\parallel$ and $a_\perp = 3/k_\perp$.[27] The results reveal that the lattice constants of HfN(001) films grown in Kr/N$_2$ discharges with continuous bias are $a_\parallel$ = 4.461 Å and $a_\perp$ = 4.580 Å, whereas with synchronized bias, $a_\parallel$ = 4.532 Å and $a_\perp$ = 4.552 Å.

Relaxed bulk HfN lattice $a_0$ constants are determined from $a_\parallel$ and $a_\perp$ as[63]

$$a_0 = a_\perp \left(1 - \frac{2\upsilon(a_\perp - a_\parallel)}{a_\perp(1 + \upsilon)}\right), \tag{6}$$

for which $\upsilon$ is the Poisson ratio. Substituting $a_\parallel$ and $a_\perp$ values, together with the experimentally-determined Poisson ratio for HfN ($\upsilon$ = 0.35)[64] into equation (5), we obtain $a_0$ = 4.517 Å for HfN(001) films grown in Kr/N$_2$ discharges with a continuous bias and $a_0$ = 4.542 Å in the synchronized-bias mode. Both are within the range of previously reported values, 4.504-4.543 Å, for polycrystalline and single-crystal HfN$_x$ layers.[35] The fact that the substrate and layer peaks for both HfN samples in Figure 7 are misaligned along $k_\parallel$ indicates the presence of in-plane-strain relaxation due to misfit dislocations.

Residual in-plane $\varepsilon_\parallel$ and out-of-plane $\varepsilon_\perp$ strains are obtained using the equations,

$$\varepsilon_\parallel = \frac{a_\parallel - a_0}{a_0} \tag{7}$$

and

$$\varepsilon_\perp = \frac{a_\perp - a_0}{a_0}, \qquad (8)$$

resulting in $\varepsilon_\parallel$ = -1.23% and $\varepsilon_\perp$ = 1.40% for HfN(001) films grown in Kr/N$_2$ discharges with continuous bias; for synchronized bias, $\varepsilon_\parallel$ = -0.21% and $\varepsilon_\perp$ = 0.22%. Thus, HfN$_x$ layers grown in Kr/N$_2$ with a continuous bias exhibit significantly larger residual compressive strain, as expected based on Kr incorporation results (Section B). The degree of in-plane relaxation at room-temperature is obtained from the relationship[14]

$$R_{lx}(300K) = \frac{a_\parallel - a_s}{a_0 - a_s} \qquad (9)$$

$R_{lx}(300K)$ is 0.81 for HfN(001) films grown in Kr/N$_2$ with continuous bias. Layers grown with synchronized bias are essentially fully relaxed with $R_{lx}(300K)$ = 0.97.

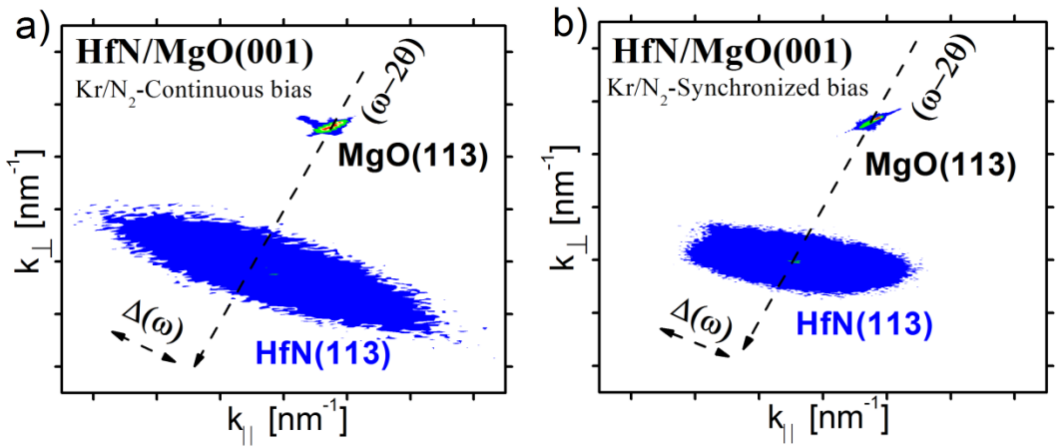

**Figure 7.** HR-RLMs acquired around asymmetric 113 reflections from 150-nm-thick HfN layers grown on MgO(001) in Kr/N$_2$ discharge at 3 mTorr. (a) The left scan is from a film deposited with a negative substrate bias $V_s$ = 100 V applied continuously during HiPIMS pulses, and (b) the right scan is from a film grown with the substrate bias synchronized to the metal-rich portion of the pulses. The average target power in both cases is 150 W.

Figure 8 is a typical bright-field XTEM image of a HfN layer grown in Kr/N$_2$ with synchronized bias revealing a sharp film/substrate interface and a flat upper film surface. The SAED pattern in the inset, obtained from the film together with part of the substrate, reveals symmetric cubic reflections. Film reflections, closer to the central transmitted beam, are aligned radially with the more intense reflections of the substrate, indicating an almost fully-relaxed lattice, in agreement with the HR-RLM results. The dark speckles in the image are due to local strain-fields associated with point-defect clusters, which are more clearly observed in the lattice-resolution image in Figure 8(b). Figure 8(a) also reveals the presence of threading dislocations extending from the film/substrate interface along the growth direction; the associated strain fields appear as dark lines in the higher-resolution image, Figure 8(c). The presence of threading dislocations is expected due to the very large film/substrate lattice mismatch, 7.46%, which gives rise to a high density of misfit dislocations.

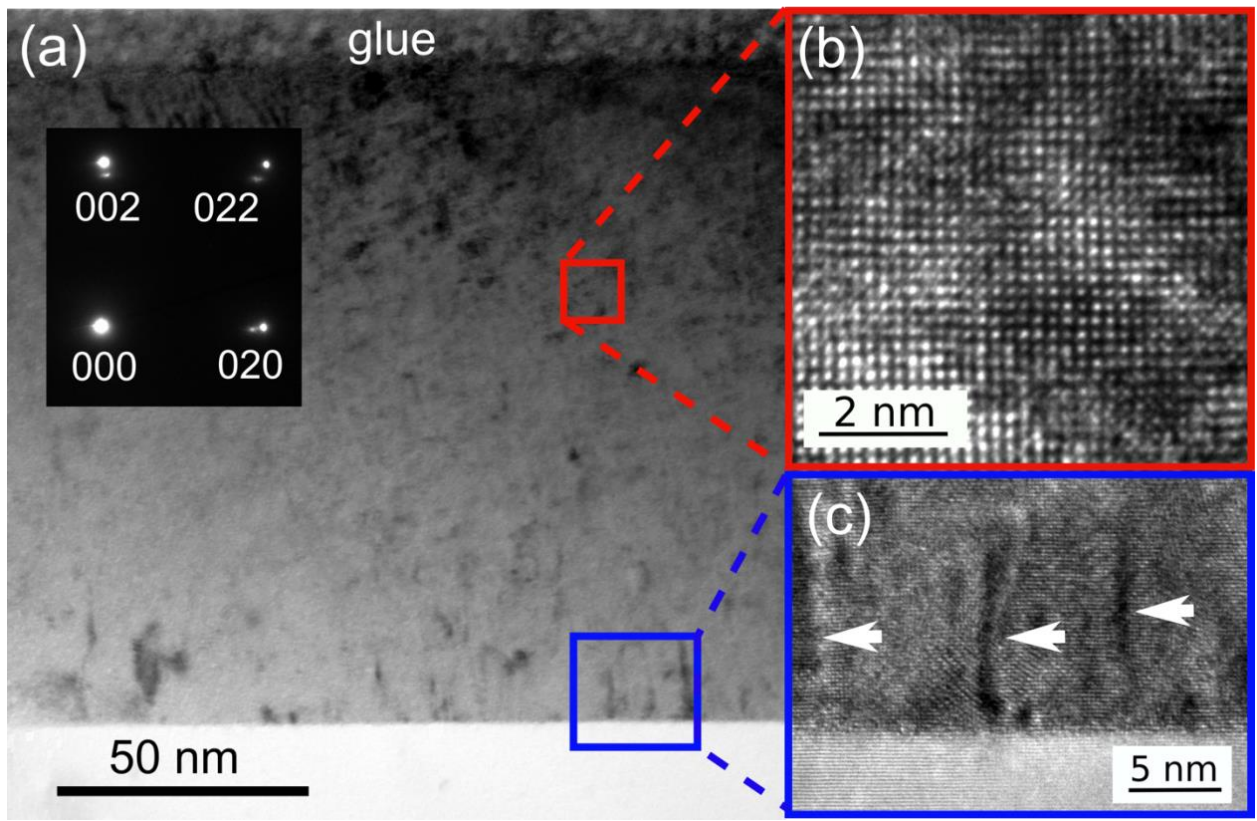

**Figure 8.** XTEM images from a 150-nm-thick HfN/MgO(001) layer grown by reactive HiPIMS in Kr/N$_2$ at 3 mTorr with $V_s$ = 100 V synchronized to the metal-rich portion of each pulse. (a) Lower-resolution bright-field XTEM micrograph with a selected-area electron-diffraction pattern shown in the inset. (b) and (c) HR-XTEM images acquired from the central region of the HfN film and the film/substrate interface, respectively. Threading dislocations in (c) are marked with white arrowheads. The average target power is 150 W.

**C. Film properties**

Room-temperature resistivities $\rho$ of epitaxial HfN(001) layers grown in Ar/N$_2$ and Kr/N$_2$ discharges with a continuous bias are 130 and 100 µΩ-cm, respectively; both significantly higher than for HfN grown in Kr/N$_2$ with synchronized bias, 70 µΩ-cm. However, all three values are much lower than those reported for polycrystalline-HfN films deposited at 400 °C for which $\rho$ ranges from 225 to 750 µΩ-cm.[65] The resistivity of stoichiometric epitaxial HfN/MgO(001) layers grown at $T_s$ = 650 °C ($T_s/T_m$ = 0.26) is 14.2 µΩ-cm.[28]

Hardness ($H$) values for HfN layers grown with a continuous bias in Ar/N$_2$ and Kr/N$_2$ discharges are 27.6±0.5 and 26.8±0.6 GPa, respectively, with elastic moduli ($E$) of 390±14 GPa for both cases. The $H$ and $E$ values of continuous-bias films are slightly higher than those grown with synchronized bias, $H$ = 25.7±0.5 and $E$ = 373±9 GPa, due to the much higher stress in the continuous-bias layers originating primarily from trapped noble-gas atoms. However, the synchronized-bias film results are significantly higher than those reported for 0.6-µm-thick polycrystalline HfN$_x$ (1.05 ≤ $x$ ≤ 1.2) layers deposited on Si(001) at 200 °C using radio-frequency sputtering, for which $H$ ranges from 22 to 21 GPa and $E$ from 285 to 370 GPa,[66] and in reasonable agreement with 0.5-µm-thick epitaxial HfN$_x$ (1.05 ≤ $x$ ≤ 1.2) layers deposited at $T_s$ = 650 °C on MgO(001) using reactive direct-current sputtering, for which $H$ ranges from 23 to 32 GPa and $E$ from 350 to 450 GPa.[35]

## IV. DISCUSSION

The results presented in section III demonstrate very-low-temperature epitaxial growth ($T_s/T_m$ < 0.10 - 0.12) of HfN on MgO(001) in the absence of applied substrate heating using reactive HiPIMS, with low-energy ion irradiation of the growing film, in both Ar/N$_2$ and Kr/N$_2$ discharges.

Mass-spectroscopy analyses show that the measured ion flux incident at the growing film during reactive HiPIMS contains both metal ions, Hf$^+$ and Hf$^{2+}$, originating from the target, and gas ions, primarily N$^+$, N$_2^+$, and either Ar$^+$ or Kr$^+$. Based on the energy-resolved measurements shown in Figure 2 and the applied substrate bias, the average incident energies for the metal ions are $E_{Hf^+}$ = 115 eV and $E_{Hf^{2+}}$ = 228 eV. For the gas ions, $E_{Ar^+}$, $E_{N_2^+}$ and $E_{Kr^+}$ = 100 eV, while $E_{N^+}$ = 112 eV. Bombardment with energetic ions and fast neutrals during HfN film growth results in overlapping shallow collision cascades which help to anneal out defects and desorb trapped noble-gas atoms. Low-energy bombardment also provides additional kinetic energy to the surface and near-surface region of the growing film, enhancing surface adatom mobilities[67] necessary for LTE.

Comparing the results for film growth with continuous vs. synchronized bias reveals that selecting the metal-rich portion of the ion flux during HiPIMS pulses is an efficient approach for improving the crystalline quality of LTE films. Sputtering Hf in Ar/$N_2$ discharges with continuous bias yields an ion intensity at the film-growth surface which is dominated by gas ions – $Ar^+$ (26.8%), $N_2^+$ (12.2%), and $N^+$ (34.6%) – and doubly-charged metal $Hf^{2+}$ ions (16.6%), with a lower contribution from singly-charged $Hf^+$ ions (9.8%). In contrast, ion irradiation conditions during sputter deposition of HfN in mixed Kr/$N_2$ discharges with continuous bias are quite different; the relative contribution of metal ions to the total ion intensity is five times larger (~50.9% for $Hf^+$ and 14.3% for $Hf^{2+}$), while the metal-charge-state ratio, $J_{Hf^{2+}}/J_{Hf^+}$, is approximately six times lower than values measured in Ar/$N_2$ discharges.

The observed changes in the composition of the ion flux between Ar/$N_2$ and Kr/$N_2$ discharges are directly related to differences in the sputtered metal-atom first and second ionization potentials compared to the first-ionization energy of the noble gas (see Table 2). The primary ionization mechanism during HiPIMS is via inelastic collisions with energetic electrons.[68,69] The upper limit of the electron-energy distribution $E_e$ is, to a first approximation, determined by the first-ionization potential of the noble gas ($IP_g^1$), which dominate the initial part of the HiPIMS pulses (Figure 3). Thus, the production rate of $Hf^{2+}$ and $Hf^+$ metal ions depends strongly on whether $IP_g^1$ is larger or smaller than $IP_{Hf}^2$ and $IP_{Hf}^1$, respectively. For Hf sputtered in Ar/$N_2$, a significant fraction of the plasma electrons have an average energy $E_e$ in the range $IP_{Hf}^2 < E_e < IP_{Ar}^1$, which is too low to ionize Ar, but high enough to produce $Hf^{2+}$ ions. The situation is different for HiPIMS discharges in Kr/$N_2$, for which ionization of Kr atoms at the beginning of the pulse tends to shift the electron-energy distribution to lower values which, in turn, results in a dramatic reduction in the $Hf^{2+}$ ion density, as shown in Figure 3. Since both $IP_{Ar}^1$ and $IP_{Kr}^1$ are much higher than $IP_{Hf}^1$, the effect on the $Hf^+$ production rate of switching noble gases is smaller. The shift in the electron-energy distribution between Ar/$N_2$ and Kr/$N_2$ discharges also explains the reduction of $N^+$ and $N_2^+$ densities in Kr/$N_2$ discharges.

Changes in ion bombardment conditions at the growing film due to switching from Ar/$N_2$ to Kr/$N_2$ are reflected in the XRD results. HfN layers grown in Kr/$N_2$ discharges exhibit a small, but clear, improvement in crystalline quality compared to films grown in Ar/N$_2$. In- and out-of-plane x-ray coherence lengths in Kr/N$_2$ layers ($\xi_\parallel$ = 28 Å; $\xi_\perp$ = 134 Å) are slightly higher (i.e., lower mosaicity) than those deposited in Ar/N$_2$ ($\xi_\parallel$ = 26 Å; $\xi_\perp$ = 114 Å). In addition, the trapped oxygen concentration in Kr/N$_2$ layers (1.8 at%) is considerably lower than for Ar/N$_2$ films (3.6 at%) indicating higher film density.

Films grown in Kr/N$_2$ discharges with a pulsed substrate bias synchronized to the metal-ion-rich portion of each HiPIMS pulse are subjected to intense Hf$^+$ bombardment (with greatly reduced noble-gas-ion irradiation) of the growing film. During the remaining portion of each pulse, $V_s$ is only ~10 eV. Based upon TRIM[43] Monte-Carlo simulations, the average Hf$^+$ ion penetration depth is estimated to be ~14 Å, with overlapping collision cascades producing shallow, near-surface dynamic atomic intermixing and noble-gas desorption, thus resulting in films with lower residual compressive stress.

The average projected ranges for Hf and N primary recoils are 10 and 8 Å due to Hf$^+$ bombardment, and 4 Å for both Hf and N recoils resulting from Kr$^+$ irradiation. Since the penetration depth for Kr$^+$ is ~6 Å, the film forming species – both the primary Hf$^+$ metal ions and the recoiled Hf lattice atoms – penetrate deeper into the near-surface region of the growing film than the inert-gas ions and can be directly incorporated into the lattice by filling residual vacancies arising due to the low growth temperature ($T_s/T_m$ < 0.10).[70] Energetic metal-ion bombardment also enhances Hf adatom mobility and, hence, leads to increased crystalline quality, due to the mass match which gives rise to more effective energy transfer. XRD $\omega$-2$\theta$ (Figure 4) and $\omega$-rocking curve (Figure 5) results both clearly show that the epitaxial Kr/N$_2$ synchronized-bias films have much lower mosaicity with a larger x-ray coherence length, by a factor of ~3×, and lower room-temperature resistivities than layers grown with continuous bias.

The reactive HiPIMS results presented here, Kr/N$_2$ vs. Ar/N$_2$ and continuous vs. synchronized bias, clearly illustrate the importance of the nature of the low-energy species bombarding the film. Selecting plasma conditions favoring bombardment by singly-ionized film-forming Hf$^+$ species results in a significant increase in the crystalline quality of LTE Hf layers. The strategy is general and can easily be transferred to the growth of other TM-nitride (as well as carbide and oxide) systems and metallic films. Each material system will require careful experimental design of HiPIMS pulse parameters (maximum target current density and duty cycle), time-dependent ion intensities and substrate bias ($V_s$ amplitude and time), and choice of noble gas for a given target material in order to obtain the maximum metal-ion/noble-gas-ion ratio and the lowest doubly- to singly-charged metal-ion ratio incident at the film-growth surface.

## IV. CONCLUSIONS

The results presented in Section III demonstrates very-low-temperature ($T_s/T_m < 0.10$) epitaxial growth of HfN thin films on MgO(001) using reactive HiPIMS with the substrate bias synchronized to the metal-ion-rich portion of the discharge. The HfN/MgO system was chosen since HfN has the highest melting point, 3310 °C,[31] of all the TM nitrides and the film/substrate lattice mismatch is large, 7.46%, both providing challenges for LTE growth.

The highest-crystalline-quality LTE HfN films are achieved in Kr/N$_2$ discharges using a substrate bias synchronized with the metal-rich portion of each HiPIMS pulse, which also decreases plasma heating such that the film growth temperature is reduced to < 70 °C ($T_s$ increases essentially linearly with deposition time to reach a maximum of value of 70 °C at 40 min into 60 min deposition runs.). The synchronized bias results in increased Hf$^+$ self-ion irradiation and reduced inert gas-ion bombardment, both leading to a significantly decreased film mosaicity, and correspondingly lower resistivity values. The layers grow with a cube-on-cube orientation, (001)$_{HfN}$∥(001)$_{MgO}$ and [100]$_{HfN}$∥[100]$_{MgO}$, with respect to the substrate, and are essentially fully relaxed. The room temperature resistivity is ~70 μΩ-cm, well below reported values for polycrystalline HfN films ($\rho$ = 225 - 750 μΩ-cm) grown at 450 °C.[65] Hardness and elastic moduli values from nanoindentation experiments are 25.7±0.5 GPa and 373±9 GPa, respectively, in good agreement with values reported for single-crystal HfN$x$ (1.05 ≤ $x$ ≤ 1.2) films grown at 650 °C ($T_s/T_m$ = 0.26) using reactive dc sputtering: 23 < $H$ < 32 GPa and 350 < $E$ < 450 GPa.[35]

Layers grown in Ar/N$_2$ discharges under continuous bias ($V_s$ = 100 V) reach a maximum substrate temperature of 120 °C ($T_s/T_m$ = 0.12) due to plasma heating and have lower crystalline quality than those grown in Kr/N$_2$ with continuous bias. This is manifested in higher mosaicity, trapped noble-gas concentration, and porosity. Switching to a Kr/N$_2$ discharge, with continuous bias, suppresses the formation of high-energy gas ions, particularly N$^+$.

## ACKNOWLEDGMENTS

This work has been supported by the Swedish Research Council (grant VR 621-2014-4882) and the Swedish Government Strategic Research Area in Materials Science on Functional materials at Linköping University (Faculty Grant SFO-Mat-LiU No. 2009-00971). We would also like to thank Merve Üstüncelik for assistance with the experimental work and Petter Larsson at Ionautics AB for technical support.

## REFERENCES

[1] U. Helmersson, S. Todorova, S.A. Barnett, J. -E. Sundgren, L.C. Markert, and J.E. Greene, J. Appl. Phys. **62**, 481 (1987). doi

[2] J.-E. Sundgren, J. Birch, G. Håkansson, L. Hultman, and U. Helmersson, Thin Solid Films **193–194**, 818 (1990). doi

[3] C.-S. Shin, D. Gall, N. Hellgren, J. Patscheider, I. Petrov, and J.E. Greene, J. Appl. Phys. **93**, 6025 (2003). doi

[4] H. Kindlund, D.G. Sangiovanni, J. Lu, J. Jensen, V. Chirita, I. Petrov, J.E. Greene, and L. Hultman, J. Vac. Sci. Technol. **32**, 030603 (2014). doi

[5] T. Lee, K. Ohmori, C.-S. Shin, D.G. Cahill, I. Petrov, and J.E. Greene, Phys. Rev. B **71**, 144106 (2005). doi

[6] P. Hedenqvist, M. Bromark, M. Olsson, S. Hogmark, and E. Bergmann, Surf. Coat. Technol. **63**, 115 (1994). doi

[7] T. Polcar, T. Kubart, R. Novák, L. Kopecký, and P. Široký, Surf. Coat. Technol. **193**, 192 (2005). doi

[8] D. McIntyre, J.E. Greene, G. Håkansson, J.-E. Sundgren, and W.-D. Münz, J. Appl. Phys. **67**, 1542 (1990). doi

[9] J.C. Sánchez-López, D. Martínez-Martínez, C. López-Cartes, A. Fernández, M. Brizuela, A. García-Luis, and J.I. Oñate, J. Vac. Sci. Technol. **23**, 681 (2005). doi

[10] L.A. Donohue, I.J. Smith, W.-D. Münz, I. Petrov, and J.E. Greene, Surf. Coat. Technol. **94–95**, 226 (1997). doi

[11] A.B. Mei, M. Tuteja, D.G. Sangiovanni, R.T. Haasch, A. Rockett, L. Hultman, I. Petrov, and J.E. Greene, J. Mater. Chem. C **4**, 7924 (2016). doi

[12] D. Gall, I. Petrov, and J.E. Greene, J. Appl. Phys. **89**, 401 (2001). doi

[13] A.B. Mei, A. Rockett, L. Hultman, I. Petrov, and J.E. Greene, J. Appl. Phys. **114**, 193708 (2013). doi

[14] A.B. Mei, B.M. Howe, C. Zhang, M. Sardela, J.N. Eckstein, L. Hultman, A. Rockett, I. Petrov, and J.E. Greene, J. Vac. Sci. Technol. A **31**, 061516 (2013). doi

[15] Q. Zheng, A.B. Mei, M. Tuteja, D.G. Sangiovanni, L. Hultman, I. Petrov, J.E. Greene, D.G. Cahill, Phys. Rev. Materials **1**, 065002 (2017). doi

[16] P. Dubey, S. Srivastava, R. Chandra, and C. V. Ramana, AIP Adv. **6**, 075211 (2016). doi

[17] H. Kindlund, D.G. Sangiovanni, L. Martínez-de-Olcoz, J. Lu, J. Jensen, J. Birch, I. Petrov, J.E. Greene, V. Chirita, and L. Hultman, APL Mater. **1**, 042104 (2013). doi

[18] S. Niyomsoan, W. Grant, D.L. Olson, and B. Mishra, Thin Solid Films **415**, 187 (2002). doi

[19] J.-S. Chun, I. Petrov, J.E. Greene, J. Appl. Phys. **86**, 3633 (1999). doi

[20] R.A. Araujo, X. Zhang, and H. Wang, J. Vac. Sci. Technol. B **26**, 1871 (2008). doi

[21] K.-H. Min, K.-C. Chun, and K.-B. Kim, J. Vac. Sci. Technol. B **14**, 3263 (1996). doi

[22] J.S. Becker and R.G. Gordon, Appl. Phys. Lett. **82**, 2239 (2003). doi

[23] H. Kim, C. Detavenier, O. van der Straten, S.M. Rossnagel, A.J. Kellock, and D.-G. Park, J. Appl. Phys. **98**, 014308 (2005). doi

[24] M. Mühlbacher, G. Greczynski, B. Sartory, N. Schalk, J. Lu, I. Petrov, J.E. Greene, L. Hultman, and C. Mitterer, Sci. Rep. **8**, 5360 (2018). doi

[25] D. McIntyre, J.E. Greene, G. Håkansson, J.-E. Sundgren, and W.-D. Münz, J. Appl. Phys. **67**, 1542 (1990). doi

[26] B.W. Karr, D.G. Cahill, I. Petrov, and J.E. Greene, Phys. Rev. B **61**, 16137 (2000). doi

[27] D. Gall, C.S. Shin, T. Spila, M. Odén, M.J.H. Senna, J.E. Greene, and I. Petrov, J. Appl. Phys. **91**, 3589 (2002). doi

28  H.S. Seo, T.Y. Lee, J.G. Wen, I. Petrov, J.E. Greene, and D. Gall, J. Appl. Phys. **96**, 878 (2004). doi

29  T.-Y. Lee, D. Gall, C.-S. Shin, N. Hellgren, I. Petrov, and J.E. Greene, J. Appl. Phys. **94**, 921 (2003). doi

30  T. Lee, H. Seo, H. Hwang, B. Howe, S. Kodambaka, J.E. Greene, and I. Petrov, Thin Solid Films **518**, 5169 (2010). doi

31  John R. Rumble, ed., CRC Handbook of Chemistry and Physics, 98th Edition (Internet Version 2018), CRC Press/Taylor & Francis, Boca Raton, FL, USA.

32  J.T. Gudmundsson, N. Brenning, D. Lundin and U. Helmersson, J. Vac. Sci. Technol. A **30**, 030801 (2012). doi

33  G. Greczynski, J. Lu, J. Jensen, S. Bolz, W. Kölker, Ch. Schiffers, O. Lemmer, J.E. Greene, and L. Hultman, Surf. Coat. Technol. **257**, 15 (2014). doi

34  G. Greczynski, J. Lu, J. Jensen, I. Petrov, J.E. Greene, S. Bolz, W. Kölker, C. Schiffers, O. Lemmer, and L. Hultman, J. Vac. Sci. Technol. A **30**, 061504 (2012). doi

35  H.S. Seo, T.Y. Lee, I. Petrov, J.E. Greene, and D. Gall, J. Appl. Phys. **97**, 083521 (2005). doi

36  J.L. Schroeder, A.S. Ingason, J. Rosén, and J. Birch, J. Cryst. Growth **420**, 22 (2015). doi

37  R.C. Powell, N.-E. Lee, Y.-W. Kim, and J.E. Greene, J. Appl. Phys. **73**, 189 (1993). doi

38  I. Petrov, P.B. Barna, L. Hultman, and J.E. Greene, J. Vac. Sci. Technol. A **21**, S117 (2003). doi

39  J. Bohlmark, M. Lattemann, J.T. Gudmundsson, A. P. Ehiasarian, Y. Aranda Gonzalvo, N. Brenning, and U. Helmersson, Thin Solid Films **515**, 1522 (2006). doi

40  G. Greczynski and L. Hultman, Vacuum **84**, 1159 (2010). doi

41  M.S. Janson, "CONTES Conversion of Time-Energy Spectra - A program for ERDA data analysis", Internal report, Uppsala University, 2004).

42  W.C. Oliver and G.M. Pharr, J. Mater. Res. **7**, 1564 (1992). doi

43  J.P. Biersack and J.F. Ziegler, in *Ion Implant. Tech.* (Springer Berlin Heidelberg, Berlin, Heidelberg, 1982), pp. 122–156.

44  R.C. Wetzel, F.A. Baiocchi, T.R. Hayes and R.S. Freund, Phys. Rev. A **35**, 559 (1987). doi

45 J.T. Gudmundsson, D. Lundin, N. Brenning, M.A. Raadu, C. Huo, and T.M. Minea, Plasma Sources Sci. Technol. **25**, 065004 (2016). doi

46 N. Brenning, J.T. Gudmundsson, M.A. Raadu, T.J. Petty, T. Minea, D. Lundin, Plasma Sources Sci. Technol. **26**, 125003 (2017). doi

47 G. Auday, P. Guillot, J. Galy and H. Brunet, J. Appl. Phys. **83**, 5917 (1998). doi

48 A. Hecimovic, K. Burcalova, and A.P. Ehiasarian, J. Phys. D. Appl. Phys. **41**, 095203 (2008). doi

49 M. Palmucci, N. Britun, T. Silva, R. Snyders, and S. Konstantinidis, J. Phys. D. Appl. Phys. **46**, 215201 (2013). doi

50 A. Ferrec. J. Keraudy, S. Jacq, F. Schuster, P.-Y. Joaun, M.A. Djouadi, Surf. Coatings. Technol. **250**, 21 (2014). doi

51 P. Sigmund, J. Vac. Sci. Technol. **17**, 396 (1980). doi

52 M.W. Thompson, Phys. Rep. **69**, 335 (1981). doi

53 P.A. Ni, C. Hornschuch, M. Panjan, and A. Anders, Appl. Phys. Lett. **101**, 224102 (2012). doi

54 D. Lundin, P. Larsson, E. Wallin, M. Lattemann, N. Brenning, and U. Helmersson, Plasma Sources Sci. Technol. **17**, 035021 (2008). doi

55 A. Hecimovic, C. Corbella, C. Maszl, W. Breilmann, and A. von Keudell, J. Appl. Phys. **121**, 171915 (2017). doi

56 T. Welzel, S. Naumov, and K. Ellmer, J. Appl. Phys. **109**, 073302 (2011). doi

57 H. Matsui, H. Toyoda, and H. Sugai, J. Vac. Sci. Technol. A **23**, 671 (2005). doi

58 G. Greczynski, I. Zhirkov, I. Petrov, J.E. Greene, and J. Rosen, J. Vac. Sci. Technol. A **36**, 020602 (2018). doi

59 D. Lundin, J.T. Gudmundsson, N. Brenning, M.A. Raadu, T.M. Minea, J. Appl. Phys. **121**, 171917 (2017). doi

60 G. Greczynski, I. Petrov, J.E. Greene, and L. Hultman, Vacuum **116**, 36 (2015). doi

61 G. Greczynski and L. Hultman, Vacuum **84**, 1159 (2010). doi

62 P. van der Sluis, J. Phys. D. Appl. Phys. **26**, A188 (1993). doi

63 C.-S. Shin, S. Rudenja, D. Gall, N. Hellgren, T.-Y. Lee, I. Petrov, and J.E. Greene, J. Appl. Phys. **95**, 356 (2004). [doi](doi)

64 A.J. Perry, V. Valvoda, and D. Rafaja, Thin Solid Films **214**, 169 (1992). [doi](doi)

65 B.O. Johansson, J. -E. Sundgren, and U. Helmersson, J. Appl. Phys. **58**, 3112 (1985). [doi](doi)

66 C. Hu, X. Zhang, Z. Gu, H. Huang, S. Zhang, X. Fan, W. Zhang, Q. Wei, and W. Zheng, Scr. Mater. **108**, 141 (2015). [doi](doi)

67 I. Petrov, P.B. Barna, L. Hultman, and J.E. Greene, J. Vac. Sci. Technol. A **21**, S117 (2003). [doi](doi)

68 C. Huo, M.A. Raadu, D. Lundin, J.T. Gudmundsson, A. Anders, and N. Brenning, Plasma Sources Sci. Technol. **21**, 045004 (2012). [doi](doi)

69 J.T. Gudmundsson, D. Lundin, N. Brenning, M.A. Raadu, C. Huo, and T.M. Minea, Plasma Sources Sci. Technol. **25**, 065004 (2016). [doi](doi)

70 G. Greczynski, J. Lu, M.P. Johansson, J. Jensen, I. Petrov, J.E. Greene, and L. Hultman, Surf. Coatings Technol. **206**, 4202 (2012). [doi](doi)